\documentclass[fleqn,twoside]{article}
\usepackage{espcrc2}

\def\beq{\begin{equation}}
\def\eeq{\end{equation}}
\def\bqry{\begin{eqnarray}}
\def\eqry{\end{eqnarray}}

\def\qbar{{\overline{q}}}
\def\sbar{{\overline{s}}}

\def\tq{{\tilde{q}}}
\def\ts{{\tilde{s}}}
\def\td{{\tilde{d}}}
\def\tu{{\tilde{u}}}
\def\tqbar{{\overline{\tq}}}

\def\tdbar{{\overline{\td}}}

\def\str{{\rm str\;}}
\def\Nhat{{\hat N}}
\def\tQ{{\tilde{Q}}}
\def\tK{{\tilde{K}}}

\newcommand{\AmS}{{\protect\the\textfont2
  A\kern-.1667em\lower.5ex\hbox{M}\kern-.125emS}}

\title{Analytic estimates of quenched penguins}

\author{Maarten Golterman\thanks{presenter at conference}\address{Dept. of Physics and Astronomy,
San Francisco State University, 1600 Holloway Ave, San Francisco,
CA 94132, USA} and
Santiago Peris\address{Grup de F{\'\i}sica Te{\`o}rica and IFAE,
Universitat Aut{\`o}noma de Barcelona, 08193 Barcelona, Spain}}

\begin{document}

\begin{abstract}
When we embed the strong penguin operator $Q_6$ in the quenched theory, it does not
remain a singlet under right-handed chiral transformations.  As a consequence, more
low-energy constants associated with this operator appear than in the unquenched theory.
We give analytic estimates of the leading constants.  The results suggest that the
effects of quenching on this operator are large.

\vspace{1pc}
\end{abstract}

\maketitle

The gluonic penguin operator \bqry Q_6=4(\sbar_L^\alpha\gamma_\mu
d_L^\beta)\sum_{q=u,d,s} (\qbar_R^\beta\gamma_\mu q_R^\alpha) \eqry ($\alpha$, $\beta$
are color indices) is one of the strong penguin operators contributing to $K\to\pi\pi$
decays, in particular to the quantity $\varepsilon'/\varepsilon$, which measures direct
CP violation. In the real world, it is a singlet under SU(3)$_R$.  However, in quenched
QCD the situation is more complicated \cite{GP1}, and the operator can be written as
\bqry
Q_6&=&\frac{1}{2}Q_6^{QS}+Q_6^{QNS}\,,\\
\frac{1}{2}Q_6^{QS}&=&
2(\sbar_L^\alpha\gamma_\mu d_L^\beta)
(\qbar_R^\beta\gamma_\mu q_R^\alpha+\tqbar_R^\beta\gamma_\mu \tq_R^\alpha)\,,
\nonumber\\
Q_6^{QNS}&=& 2(\sbar_L^\alpha\gamma_\mu d_L^\beta)
(\qbar_R^\beta\gamma_\mu q_R^\alpha-\tqbar_R^\beta\gamma_\mu \tq_R^\alpha)\,,
\nonumber
\eqry
where $q$ is summed over $u,d,s$, and
$\tq=\tu,\td,\ts$ are the bosonic (ghost) quarks
used to define the quenched theory \cite{morel}.  We see that while
$Q_6^{QS}$ is a singlet under the quenched flavor
group SU(3$|$3)$_R$ \cite{BG}, $Q_6^{QNS}$ is not.

To leading order in chiral perturbation theory (ChPT),
these operators maybe represented
by \cite{CBetal,GP1}
\bqry
Q_6^{QS}&\to& -\alpha^{(8,1)}_{q1}\str(\Lambda L_\mu L_\mu)\\&&
+\alpha^{(8,1)}_{q2}\str(2B_0\Lambda(\Sigma M+M\Sigma^\dagger))\,,
\nonumber\\
Q_6^{QNS}&\to& f^2\alpha^{NS}_q\str(\Lambda\Sigma\Nhat\Sigma^\dagger)\,,
\nonumber\\
\Nhat&=&\frac{1}{2}\,{\rm diag}(1,1,1,-1,-1,-1)\,, \nonumber \nonumber \eqry with
$\qbar\Lambda q=\sbar d$, $M$ the quark-mass matrix, and $\str$ is the supertrace.
$\Nhat$ represents the non-singlet structure of $Q_6^{QNS}$. The low-energy constants
(LECs) $\alpha^{(8,1)}_{q1,2}$ correspond to LECs also appearing in the unquenched
theory, whereas the new LEC $\alpha^{NS}_q$ is a quenched artifact. Note that
$\alpha^{NS}_q$ is of order $p^0$, while the others are of order $p^2$.\footnote{See
Ref.~\cite{GP1} for more discussion of this observation.}

The new LEC $\alpha^{NS}_q$ shows up for instance in 
\bqry \langle 0|Q_6|K^0\rangle
\!\!\!&=&\!\!\!\frac{4i}{f}\Biggl\{\left(\frac{1}{2}
\alpha^{(8,1)}_{q2}+\beta^{NS}_{q}\right)
(M_K^2-M_\pi^2)\nonumber\\
&&\hspace{-2.2cm}
+\frac{\alpha^{NS}_q}{16\pi^2}
\sum_{q=u,d,s}\left(M_{sq}^2\log{M_{sq}^2}-M_{dq}^2\log{M_{dq}^2}
\right)\Biggr\}\,,\nonumber\\
M^2_{qq'}&=&B_0(m_q+m_{q'})\,,
\eqry
where $\beta^{NS}$ is a higher-order LEC associated with $Q_6^{QNS}$.
The appearance of $\alpha^{NS}_q$ is important because it
influences the determination of $\alpha^{(8,1)}_{q2}$ from a simulation,
but it is hard to determine $\alpha^{NS}_q$ this way in practice.

However, there is a trick which makes $\alpha^{NS}_q$ more easily
accessible \cite{GP2}.  One rotates $d_L\to\td_L$ in $Q_6^{QNS}$
(calling this $\tQ_6^{QNS}$), and considers $\tK^0\to 0$ with $\tK^0$
made out of an anti-$s$ quark and a ghost-$d$ quark.  One finds
that
\bqry
\langle 0|\tQ_6^{QNS}|\tK^0\rangle&=&
2if\alpha^{NS}_q+O(p^2)\,.
\eqry

We may now estimate $\alpha^{NS}_q$ analytically in the following
way \cite{MGSP}.  First we fierz
\bqry
\tQ_6^{QNS}&\!\!\!\!=\!\!\!\!&
-4\left((\sbar P_R q)(\qbar P_L\td)\!+\!(\sbar P_R\tq)(\tqbar P_L\td)\right)
\eqry
(taking into account that the ghost fields commute),
and then Wick contract to find
\bqry
\langle 0|\tQ_6^{QNS}|\tK^0\rangle&=&
(\langle\sbar s\rangle
-\langle\tdbar\td\rangle)\langle 0|\sbar\gamma_5\td|\tK^0\rangle
\\
&&\hspace{-2cm}-4\langle 0|(\sbar P_R\,
\underline{q\qbar}\,P_L\td+\sbar P_R\,
\underline{\tq\tqbar}\,P_L\td)|\tK^0\rangle\,,
\nonumber
\eqry
correct to order $1/N_c^2$.  It can be shown \cite{MGSP}
that the unfactorized term
(2nd line) is order $p^2$ in ChPT, and thus it does not contribute to
$\alpha^{NS}_q$.  Using that
\bqry
\langle\tdbar\td\rangle=
-\langle\sbar s\rangle=\frac{1}{2}f^2B_0\nonumber
\eqry
(note the minus sign: ghost quarks commute!), one obtains
\bqry
\alpha^{NS}_q&=&-\frac{1}{2}f^2B_0^2
\left(1+O\left(\frac{1}{N_c^2}\right)\right)\,.
\eqry

We now consider the singlet operator $Q_6^{QS}$, which is interesting because it will
yield a result for $\alpha^{(8,1)}_{q1}$, which we may then compare to its unquenched
value, as well as to our estimate of $\alpha^{NS}_q$.  $Q_6^{QS}$ can be fierzed into
\bqry Q_6^{QS}&\!\!\!\!=\!\!\!\!&-8\left((\sbar P_R q)(\qbar P_L d) -(\sbar P_R
\tq)(\tqbar P_L d)\right) . \eqry Since the quark and ghost-quark propagators are equal
(by construction!), contributions in which $q$ and $\qbar$ or $\tq$ and $\tqbar$ are
contracted cancel.  Thus, to order $1/N_c^2$ {\it only} the factorized contribution
survives, and leads to the long-known result (with operator mixing taken into account to
leading-log order) \bqry \alpha^{(8,1)}_{q1}&=&-8L_5 f^2 B_0^2
\left(1+O\left(\frac{1}{N_c^2}\right)\right)\,, \eqry where $L_5$ is one of the
Gasser--Leutwyler $O(p^4)$ LECs of the strong effective lagrangian \cite{GL}.

{}From these results, we draw the following two conclusions.
First, $\alpha^{NS}_q$ {\it is not small compared to}
$\alpha^{(8,1)}_{q1}$:
\bqry
\frac{\alpha^{NS}_q}{\alpha^{(8,1)}_{q1}}
&=&\frac{1}{16L_5}\sim 60\,,
\eqry
using that $L_5\sim 10^{-3}$ (in quenched \cite{alpha,BET} and
unquenched \cite{GL} QCD).\\
$\bullet$
$\alpha^{NS}_q$ {\it can thus not be ignored in any quenched computation
which involves} $Q_6^{QNS}$.

In the unquenched case, it was found that the {\it unfactorized}
contribution to $\alpha^{(8,1)}_{q1}$
maybe {\it large}, and of the same sign as the factorized contribution
\cite{HPdR}.  Since the unfactorized contribution vanishes in the
quenched case, \\
$\bullet$ {\it the quenched value of}
$\alpha^{(8,1)}_1$ {\it maybe substantially smaller than
the unquenched one.}

We close with a few remarks.

First, these estimates maybe extended to the partially quenched case,
in which $N$ sea quarks are added to the quenched theory, see
Ref.~\cite{MGSP}.

Second, to extract leading order LECs,
one works in the chiral limit.
While the quenched theory is probably singular in the chiral limit,
we believe that this is not a problem for the calculation of
$\alpha^{NS}_q$ and $\alpha^{(8,1)}_{q1}$.

Finally, we believe that it should be instructive to adapt analytic
estimates of other
hadronic quantities to the quenched and partially quenched cases.
This will give {\it quantitative} information about the effects of
quenching, which ChPT by itself cannot provide.

\section*{Acknowledgements}

MG was supported in part by the US Dept.
of Energy, and SP was supported by
CICYT-FEDER-FPA2002-00748, 2001 SGR00188 and by TMR EC-Contracts
HPRN-CT-2002-00311 (EURIDICE).


\begin{thebibliography}{9}

\bibitem{GP1}
M.~Golterman and E.~Pallante,
%``Effects of quenching and partial quenching on penguin matrix elements,''
JHEP {\bf 0110}, 037 (2001)
[arXiv:hep-lat/0108010].
%%CITATION = HEP-LAT 0108010;%%

\bibitem{morel}
A.~Morel,
%``Chiral Logarithms In Quenched QCD,''
J.\ Phys.\ (France) {\bf 48}, 1111 (1987).
%%CITATION = JOPQA,48,1111;%%

\bibitem{BG}
C.~W.~Bernard and M.~F.~Golterman,
%``Chiral perturbation theory for the quenched approximation of QCD,''
Phys.\ Rev.\ D {\bf 46}, 853 (1992)
[arXiv:hep-lat/9204007].
%%CITATION = HEP-LAT 9204007;%%

\bibitem{CBetal}
C.~W.~Bernard, T.~Draper, A.~Soni, H.~D.~Politzer and M.~B.~Wise,
%``Application Of Chiral Perturbation Theory To K $\to$ 2 Pi Decays,''
Phys.\ Rev.\ D {\bf 32}, 2343 (1985).
%%CITATION = PHRVA,D32,2343;%%

\bibitem{GP2}
M.~Golterman and E.~Pallante,
%``On the effects of (partial) quenching on penguin contributions to  K $\to$ pi pi,''
arXiv:hep-lat/0212008.
%%CITATION = HEP-LAT 0212008;%%

\bibitem{MGSP}
M.~Golterman and S.~Peris,
%``Analytic estimates for penguin operators in quenched QCD,''
arXiv:hep-lat/0306028.
%%CITATION = HEP-LAT 0306028;%%

\bibitem{GL}
J.~Gasser and H.~Leutwyler,
%``Chiral Perturbation Theory: Expansions In The Mass Of The Strange Quark,''
Nucl.\ Phys.\ B {\bf 250}, 465 (1985).
%%CITATION = NUPHA,B250,465;%%

\bibitem{alpha}
J.~Heitger, R.~Sommer and H.~Wittig  [ALPHA Collaboration],
%``Effective chiral Lagrangians and lattice QCD,''
Nucl.\ Phys.\ B {\bf 588}, 377 (2000)
[arXiv:hep-lat/0006026].
%%CITATION = HEP-LAT 0006026;%%

\bibitem{BET}
W.~A.~Bardeen, E.~Eichten and H.~Thacker,
%``Low energy chiral Lagrangian parameters for scalar and pseudoscalar  mesons,''
arXiv:hep-lat/0209164.
%%CITATION = HEP-LAT 0209164;%%

\bibitem{HPdR}
T.~Hambye, S.~Peris and E.~de Rafael,
%``Delta(I) = 1/2 and epsilon'/epsilon in large-N(c) QCD,''
JHEP {\bf 0305}, 027 (2003)
[arXiv:hep-ph/0305104].
%%CITATION = HEP-PH 0305104;%%

\end{thebibliography}
\end{document}